\documentclass[twocolumn]{revtex4}

\def\bea{\begin{eqnarray}}
\def\eea{\end{eqnarray}}
\def\bec{\begin{center}}
\def\ec{\end{center}}

\def\beq{\begin{equation}}
\def\eeq{\end{equation}}

\def\f{\frac}

\def\plb{Phys. Lett. B}

\def\f#1#2{\frac{#1}{#2}}

\begin{document}
\draft
\tighten
\preprint{KAIST-TH 03/??}
\title{\large \bf Flavor Structure of
Scherk-Schwarz Supersymmetry Breaking 
\\for
Quasi-Localized Matter Fields}
\author{Kiwoon Choi\footnote{kchoi@hep.kaist.ac.kr}}
\address{Department of Physics, Korea Advanced Institute of Science and Technology\\ Daejeon
305-701, Korea}
\date{\today}
\begin{abstract}
We discuss the flavor structure of 
soft supersymmetry breaking parameters 
in 5-dimensional orbifold field theories
in which $N=1$ supersymmetry is broken by the 
Scherk-Schwarz boundary condition and
hierarchical 4-dimensional Yukawa couplings are obtained by
quasi-localizing the matter fields in extra dimension.
In such theories, the resulting
soft scalar masses and trilinear scalar couplings
at the compactification scale are highly flavor-dependent,
but appropriately suppressed in correlation
with Yukawa couplings.
Those flavor violations  
can give interesting phenomenological consequences
at low energies as well as constrain
the mechanism of Yukawa coupling generation.

\end{abstract}
\maketitle

\section{Introduction}

Supersymmetry (SUSY) is one of the prime candidates for
new physics beyond the standard model \cite{nilles}.
An important issue in supersymmetric theories is to
understand how SUSY is broken in low energy
world. It has been known that theories with compact extra dimension
provide an attractve way to break SUSY,
imposing nontrivial boundary conditions on the field variables.
This mechanism which has been proposed originally by Scherk and Schwarz (SS)
\cite{ss}
can be interpreted as a spontaneous SUSY breaking 
induced by the auxiliary component of higher dimensional supergravity
(SUGRA) multiplet \cite{ss1}. 
Extra dimension can provide
also an attractive mechanism to 
generate hierarchical Yukawa
couplings \cite{yukawa}.
The quark and lepton fields can be quasi-localized
in extra dimension in a natural manner, and then 
their 4-dimensional (4D) Yukawa couplings are determined by
the wavefunction overlap factor 
$e^{-M\pi R}$ where
$M$ is a combination of mass parameters in higher dimensional
theory and $R$ is the length of extra dimension.
This allows that hierarchical Yukawa couplings are obtained from
fundamental mass parameters having the same order of magnitude.
In this talk, we discuss the flavor structure of soft SUSY
breaking parameters induced by the SS boundary condition
imposed on the matter fields which are 
quasi-localized  to generate hierarchical Yukawa
couplings \cite{choi,choi2}.

\medskip

\section{
yukawa couplings and
soft parameters of quasi-localized matter fields}

To proceed, let us consider a generic 5D gauge theory
coupled to the minimal 5D SUGRA on
$S^1/Z_2$. The action of the model is given by
\cite{5dsugra} 
\bea \label{5daction1}
&&\int d^5x\sqrt{-G} \left[\,\left(\, \frac{1}{2}{\cal
R}+\frac{1}{2}\bar{\Psi}^i_M\gamma^{MNP}D_N\Psi_{iP}
\right.\right.\nonumber \\
&&\left.-\frac{3}{4}C_{MN}C^{MN}
\,\right)+\frac{1}{{g}_{5a}^2}
\left(-\frac{1}{4}F^{aMN}F^a_{MN}\right.\nonumber \\
&&\left.+\frac{1}{2}D_M\phi^aD^M\phi^a
+\frac{i}{2}\bar{\lambda}^{ai}\gamma^MD_M\lambda^a_i
\right)\nonumber \\
&&\left.+|D_Mh_I^i|^2+i\bar{\Psi}_I\gamma^MD_M\Psi_I+
i\tilde{m}_I\epsilon(y)\bar{\Psi}_I\Psi_I\, \right]
\eea
where ${\cal R}$ is the Ricci scalar of the 5D metric
$G_{MN}$, $\Psi^i_M$ ($i=1,2$) are
the symplectic Majorana gravitinos,
$C_{MN}=\partial_MB_N-\partial_NB_M$ is the graviphoton field strength, and
$y$ is the 5th coordinate with a fundamental range
$0\leq y\leq\pi$.
Here $\phi^a,A_M^a$ and $\lambda^{ia}$ are 5D scalar, vector and
symplectic Majorana
spinors constituting a 5D vector multiplet,
$h_I^i$ and $\Psi_I$  are 5D scalar and Dirac spinor constituting
the $I$-th hypermultiplet
with kink mass $\tilde{m}_I\epsilon(y)$.
The kink masses are related to the gauging of graviphoton
as indicated by the following covariant derivatives of
hypermultiplets:
\bea
\label{gauging}
&& D_M h^i_I=\left(\partial_M+i\tilde{m}_I\epsilon(y)B_M \right)h^i_I+...,
\nonumber \\
&& D_M\Psi_I=\left(\partial_M+i\tilde{m}_I\epsilon(y)B_M\right)\Psi_I+...,
\nonumber
\eea
where the ellipses stand for other gauge interactions.
Although not required within 5D SUGRA, it is not
unreasonable to assume that the kink masses
are {\it quantized} in an appropriate
unit, which will be
adopted here.

It is convenient to write the  5D action (\ref{5daction1}) in $N=1$
superspace \cite{Arkani-Hamed:2001tb}.
For the 5D SUGRA multiplet, we keep only the radion superfield
$$
T=R+iB_5+\theta \Psi_{5R}+
\theta^2F^T\,,
$$
where $R=\sqrt{G_{55}}$ denotes the radius of the compactified
5-th dimension, and
$\Psi_{5R}=
\frac{1}{2}(1+\gamma_5)\Psi^{i=2}_{M=5}$.
For 5D vector multiplets and hypermultiplets,
the relevant piece of the action is given by \cite{Arkani-Hamed:2001tb}
\bea \label{5daction2}
&&\int d^5x \,\left[\,\int d^4\theta
\,\frac{T+T^*}{2}\left(
e^{-\tilde{m}_I(T+T^*)|y|}H_I^*H_I
\right.\right.
\nonumber \\
&&\quad\quad\quad
\quad\quad\quad\left.+e^{\tilde{m}_I(T+T^*)|y|}H^{c*}_IH^{c}_I\right)
\nonumber \\ && \quad\quad\quad\left. +\left\{\,
\int d^2\theta\, \frac{1}{4{g}_{5a}^2}TW^{a\alpha}W^a_{\alpha}
+{\rm h.c.}\,\right\} \right]\,,\eea 
where $W^{a}_{\alpha}$ is the
chiral spinor superfield for the 5D vector superfield 
$$
{\cal V}^a=-\bar{\theta}\sigma^\mu\theta A^a_\mu
-i\bar{\theta}^2\theta\lambda^a+i\theta^2\bar{\theta}\bar{\lambda}^a
+\frac{1}{2}\theta^2\bar{\theta}^2D^a\,,
$$
and
\bea
\label{hyper}
&&H_I=e^{\tilde{m}_IT|y|}(h^1_I+\theta \psi_I+\theta^2 F_I)\,,
 \nonumber \\
&&H_I^c=e^{-\tilde{m}_IT|y|}(h^{2*}_I+
\theta\psi^c_I+\theta^2 F^c_I)\,,
\eea
where
$\lambda^a=\frac{1}{2}(1-\gamma_5)\lambda^{a1}$,
$\psi_I=\frac{1}{2}(1-\gamma_5)\Psi_I$,
and $\bar{\psi^c}_I=\frac{1}{2}(1+\gamma_5)\Psi_I$.

As the theory is
orbifolded by $Z_2: y\rightarrow -y$, all 5D fields should have a
definite boundary condition under $Z_2$.
To give a massless 4D zero mode, the vector superfield 
${\cal V}^a$ is required to be $Z_2$-even,
while the hypermultiplet can have any $Z_2$-boundary
condition:
\bea
{\cal V}^a(-y)&=&{\cal V}^a(y)\,,
\nonumber \\
H_I (-y) &=& z_I H_I (y)\,,
\nonumber \\
H_I^c(-y) &=& -z_I H_I (y),
\nonumber
\eea
where $z_I=\pm 1$.
In the superfield basis of $(H_I, H^c_I)$, 
the wavefunctions of 4D zero modes 
are $y$-independent constant.
However in the original 5D field basis for which the 5D action
is given by (\ref{5daction1}), the wavefunction
of $Q_I$ is given by
$e^{-z_I\tilde{m}_IT|y|}$,
so $Q_I$ is quasi-localized at $y=0$ if 
$z_I\tilde{m}_I>0$,
and at $y=\pi$ if $z_I\tilde{m}_I<0$.
Such quasi-localization of matter zero modes
can generate hierarchical Yukawa couplings in a natural manner
\cite{yukawa}.

In addition to the bulk action (\ref{5daction2}),
there can be brane actions at the fixed points
$y=0,\pi$.
The general covariance requires that
the 4D metric in brane action should be the
4D component of the 5D metric at the fixed point, i.e.
$G_{\mu\nu}|_{y=0,\pi}$ ($\mu,\nu=0,1,2,3$).
Using the general covariance and also
the $T$-dependent field redefinitions
(\ref{hyper}),
one can easily find
the $T$-dependence of brane actions \cite{Arkani-Hamed:2001tb}.
For instance,
the brane actions which would be relevant for
Yukawa couplings and soft parameters
are given by
\bea
\label{braneaction}
&&\int\,d^5x\,
\left[\,\,\delta(y)\,\left\{
\int d^4\theta \,L_{I\bar{J}}\Phi_I\Phi^*_J
\right.\right.\nonumber \\
&& \left.+\left(\,\int d^2\theta
\,\lambda_{IJK}\Phi_I\Phi_J\Phi_K
+\frac{1}{4} \omega_a W^{a\alpha}W^a_{\alpha}
+{\rm h.c.}\,\right)\,\right\}
\nonumber \\
&&+\,\,\delta(y-\pi)\left\{
\int d^4\theta \, e^{-(m_I\pi T+m_J\pi T^*)}
L^\prime_{I\bar{J}}\Phi_I\Phi^{*}_J
\right.
\\
&&+\left(\,\int d^2\theta \,
e^{-(m_I+m_J+m_K)\pi T}\lambda^\prime_{IJK}
\Phi_I\Phi_J\Phi_K
\right.
\nonumber \\
&&\left.\left.\left.
+\frac{1}{4}
\omega_a^\prime
W^{a\alpha}W^a_{\alpha}\,\right)+{\rm h.c.}\,\right\}\,\right]
\,, \nonumber 
\eea
where $\Phi_I=H_I$ ($z_I=1$)
or $\Phi_I=H_I^c$ ($z_I=-1$), and
$$
m_I\equiv z_I\tilde{m}_I.
$$
Here $L_{I\bar{J}}$ ($L_{I\bar{J}}^\prime$) are generic hermitian functions
of the brane superfield $Z$ and $Z^*$
at $y=0$ ($Z^\prime$ and $Z^{\prime *}$ at $y=\pi$),
and $\omega_a$ and $\lambda_{IJK}$
($\omega^\prime$ and $\lambda^\prime_{IJK}$) are
generic holomorphic functions of $Z$ ($Z^\prime$).

With compact extra dimension,
SUSY can be broken by
imposing nontrivial boundary conditions on the field variables.
Such SUSY breaking has been proposed 
originally by Scherk and Schwarz \cite{ss}, and
can be interpreted as a spontaneous SUSY breaking 
induced by the auxiliary component of higher dimensional 
SUGRA multiplet \cite{ss1}.
In 5D orbifold field theory,
the SS  boundary condition is given by
\bea
&&\lambda^{ia}(y+2\pi)=\left(e^{2\pi i\vec{\omega}\cdot\vec{\sigma}}
\right)^i_j
\lambda^{aj}(y)\,,\nonumber\\
&& h^i_I(y+2\pi)=\left(e^{2\pi i\vec{\omega}\cdot\vec{\sigma}}\right)^i_j
h^j_I(y)\,,
\eea
where 
$\lambda^{ia}$ and $h^i_I$ ($i=1,2$) are the $SU(2)_R$ doublet
gauginos and hypermultiplet scalars, respectively, and
$\vec{\omega}=(\omega_1,\omega_2,0)$.
It has been noticed that 
imposing the SS  boundary condition
is equivalent to turning on the $F$-component of
the radion superfield as \cite{ss1} 
\beq
\label{ss=radion}
F^T=2R(\omega_2-i\omega_1)\,.
\eeq
A small but nonzero value of $F^T$ can be 
achieved dynamically when a proper radion stabilization mechanism
is introduced \cite{radion}.

With the identification (\ref{ss=radion}),
SUSY breaking soft parameters
induced by the SS boundary condition can be
most easily computed by constructing the effective action
of 4D matter superfields and the radion superfield 
in $N=1$ superspace.
Let $V^a$ denote the massless 4D vector superfield
originating from
${\cal V}^a$, and
$Q_I$ to be the massless 4D chiral superfield 
originating from $H_I$ ($z_I=1$) or $H_I^c$ ($z_I=
-1$). 
Their 4D effective action can be written as
\bea
\label{4deffective1}
&&\quad\quad\quad\quad
\left[\,\int d^4 \theta\,\, Y_{I\bar{J}}Q_IQ^*_J\,\right]
\nonumber \\
&&+\left[\,\int d^2 \theta
\,\left(\f{1}{4} f_a W^{a\alpha} W^a_\alpha + \tilde{y}_{IJK}Q_IQ_JQ_K\,\right)
\,
\right],
\nonumber
\eea
where $Y_{I\bar{J}}$ are hermitian wave function coefficients,
$f_a$ are holomorphic gauge kinetic functions,
and $\tilde{y}_{IJK}$ are holomorphic Yukawa couplings.
Using (\ref{5daction2}) and (\ref{braneaction}),
it is straightforward to find
\bea
\label{4deffective}
Y_{I\bar{J}}&=&Y_I\delta_{IJ}
+L_{I\bar{J}}(Z,Z^*)
+\frac{L^\prime_{I\bar{J}}(Z^\prime,Z^{\prime *})
}{e^{(m_I\pi T+m_J\pi T^*)}}\,,
\nonumber \\
f_a&=&\frac{2\pi}{g_{5a}^2}T+\omega_a(Z)+\omega^\prime_a(Z^\prime)\,,
\nonumber \\
\tilde{y}_{IJK}&=&\lambda_{IJK}(Z)+\frac{\lambda^\prime_{IJK}(Z^\prime)}{
e^{(m_I+m_J+m_K)\pi T}}\,, \eea 
where 
$$
Y_I=\frac{\Lambda}{m_I}\left( 1- e^{-m_I\pi (T+T^*)}
\right)
$$
for the cutoff scale $\Lambda$ of
5D orbifold field theory.
Note that 5D SUSY 
enforces that the Yukawa couplings of $Q_I$ originate entirely
from the brane action (\ref{braneaction}).

A naive dimensional analysis in the large radius limit
$R\gg 1/\Lambda$
suggests that
\bea
g_{5a}^2&=&{\cal O}(\pi R )\,,
\nonumber \\
L_{IJ},L'_{IJ}&=&{\cal O}(1)
\,,
\nonumber \\
\omega_a,\omega'_a&=&{\cal O}\left(
1/\pi R\Lambda\right)
\nonumber \\
\lambda_{IJK},\lambda_{IJK}'&=&{\cal O}\left(\sqrt{
\pi R\Lambda}\right)\,.
\nonumber
\eea
This shows that the brane gauge kinetic functions $\omega_a,\omega'_a$ are 
suppressed by $1/\pi R\Lambda$ compared to the 
bulk gauge kinetic functions.
If $|m_I|<1/R$, so the matter zero mode $Q_I$ is equally
spread over the 5-th dimension,
the brane wavefunction coefficients $L_{I\bar{J}}$
and $L_{I\bar{J}}'$ are similarly supressed by $1/\pi R\Lambda$
compared to the bulk wavefunction coefficients $Y_I$.
On the other hand, if $|m_I|> 1/R$, so $Q_I$ is quasi-localized,
the brane wavefunction coefficients appear to be less suppressed
since $L_{I\bar{J}}/Y_I,L_{I\bar{J}}/Y_I={\cal O}(m_I/\Lambda)$.
In the following, 
we will ignore all the brane wavefunction coefficients and brane
gauge kinetic
functions under the assumption that
$|m_I|/\Lambda$ are small enough.
In fact, a phenomenologically favored parameter region
is given by \cite{choi2}
\bea
\frac{1}{\pi R\Lambda}&=&{\cal O}(10^{-2}-10^{-3})\,,
\nonumber \\
\frac{m_I}{\Lambda}&=&{\cal O}(10^{-1}-10^{-2})\,,
\nonumber 
\eea
which would justify our assumption.

Let $y_{IJK}$, $M_a$, $m^2_{I\bar{J}}$, and $A_{IJK}$
denote the Yukawa couplings, gaugino masses,
soft scalar masses, trilinear scalar couplings, 
respectively, 
for the {\it canonically normalized}
matter superfields $Q_I=\phi^I+\theta \psi^I+\theta^2 F^I$
and gauginos $\lambda^a$ which are renormalized at the
compactification scale $M_{KK}$:
\bea
&&\frac{1}{2}y_{IJK}\phi_I\psi_J\psi_K-
\frac{1}{2}M_a\lambda^a\lambda^a-
\frac{1}{2}m^2_{I\bar{J}}\phi^I\phi^{J*}
\nonumber \\
&&\quad\quad\quad
-\frac{1}{6}A_{IJK}\phi^I\phi^J\phi^K+{\rm h.c.}
\nonumber
\eea
We then find (in the unit with $\Lambda=1$) \cite{soft}
\bea
\label{yukawa}
&&y_{IJK}=\frac{1}{\sqrt{
Y_IY_JY_K}}
\left(\lambda_{IJK}+\frac{\lambda^\prime_{IJK}}{e^{(m_I+m_J+m_K)\pi kT}}
\right),
\nonumber \\
&&m^2_{I\bar{J}}= 
\delta_{IJ}
\left(
\frac{2\pi m_IR}{e^{m_I\pi (T+T^*)/2}-
e^{-m_I\pi (T+T^*)/2}}\right)^2
\left|\frac{F^T}{2R}\right|^2,
\nonumber \\
&&\frac{A_{IJK}}{y_{IJK}}=-F^T
\frac{\partial}{\partial T}\ln
\left(\frac{\lambda_{IJK}+\lambda^\prime_{IJK}e^{-(q_I+q_J+q_K)\pi kT}}{
Y_IY_JY_K}\right),
\nonumber \\
&&M_a=\frac{F^T}{2R}.
\eea

As we have noticed, $Q_I$ with $m_I<0$ is quasi-localized at $y=\pi$
with an exponentially small wavefunction
($e^{m_I\pi R}$)  at $y=0$,
while $Q_I$ with $m_I>0$ is quasi-localized at $y=0$.
As a result, in the case that Yukawa couplings
originate from  $y=0$, 
the quark/lepton superfields with $m_I<0$ would have 
(exponentially) small canonical Yukawa couplings, while the 
quark/lepton superfields
with $m_I>0$ can have Yukawa couplings of order unity.
Then one can obtain 
hierarchical Yukawa couplings 
with an appropriate set of (quantized) kink masses
having the same order of magnitude.
The results of (\ref{yukawa}) show that
the trilinear $A$-coefficients $A_{IJK}$ at $M_{KK}$
 induced by the SS
boundary condition
are essentially of the
order of $y_{IJK}M_a$, however the ratios $A_{IJK}/y_{IJK}$
are {\it not} universal.
Such non-universal $A_{IJK}/y_{IJK}$
can lead to interesting flavor violations
of $LR/RL$-type at the weak scale.
The squark/slepton masses at $M_{KK}$ induced by the 
SS boundary condition
are {\it not}
universal also, but they are (exponentially) suppressed
for quasi-localized quark/lepton superfields.
Still those non-universal squark/slepton masses at $M_{KK}$
can cause  interesting flavor-violations of $LL/RR$-type
at the weak scale.

\medskip

\section{a simple model}

The results of (\ref{yukawa})
show that the soft parameters
at the compactification scale induced by the SS
boundary condition for quasi-localized quark/lepton superfields
are highly  {\it flavor dependent}.
The resulting flavor-violations are suppressed in parallel to
the suppressed Yukawa couplings, however 
still they can give interesting flavor-violating processes
at the weak scale.
To be more concrete, let us consider
the case that the Yukawa couplings come from
the brane action at $y=0$ and the Higgs superfields are
brane fields confined at $y=0$.
In this case, the  Yukawa couplings and soft parameters
at $M_{KK}$ can be written as \cite{choi,choi2}
\bea
\label{radiondomination}
y_{IJ}&=&\frac{\lambda_{IJ}\ln(1/\epsilon)}{\pi R}
\sqrt{\frac{N_IN_J}{(1-\epsilon^{2N_I})(
1-\epsilon^{2N_J})}},
\nonumber \\
M_a&=& \frac{F^T}{2R},\nonumber \\
A_{IJ}&=&2y_{IJ}\ln(1/\epsilon)\left(
\frac{N_I}{\epsilon^{-2N_I}-1}+
\frac{N_J}{\epsilon^{-2N_J}-1}
\right)\frac{F^T}{2R},
\nonumber \\
m^2_{I\bar{J}}&=&\delta_{IJ}\left(\,
2\ln(1/\epsilon)\frac{N_I}{\epsilon^{N_I}-
\epsilon^{-N_I}}\left|\frac{F^T}{2R}\right|\,\right)^2,
\eea
where $y_{IJ}=y_{IJK}$, $\lambda_{IJ}=\lambda_{IJK}$
with the last subscript standing for the Higgs field
confined at $y=0$,
$\epsilon\approx 0.2$ is the Cabbibo angle,
 and
\beq
N_I=\frac{m_I\pi R}{
\ln(1/\epsilon)}
\eeq
for the kink mass $m_I=z_I\tilde{m}_I$ 
of the 5D hypermultiplet ($H_I, H^c_I$)
whose zero mode corresponds to the $I$-th quark/lepton superfield.
Here we will assume that
$m_I$ are quantized in such a way that
$N_I$ are integers.

Let $\psi_I=\{q_i, u_i, d_i, \ell_i, e_i\}$ 
($i=1,2,3$) denote the
known three generations of
the left-handed quark-doublets ($q_i$), up-type 
antiquark-singlets ($u_i$), down-type antiquark singlets
($d_i$), lepton-doublets ($\ell_i$), and anti-lepton singlets 
($e_i$).  The Yukawa 
couplings can be written as
\bea
{\cal L}_{\rm Yukawa}=
 y^u_{ij}H_2q_iu_j+y^d_{ij}H_1q_id_j+y^\ell_{ij}H_1\ell_ie_j
\nonumber
\eea
and the squark/sleptons $\phi_I=\{\tilde{q}_i,\tilde{u}_i, \tilde{d}_i,
\tilde{\ell}_i, \tilde{e}_i\}$ 
have the soft SUSY breaking couplings:
\bea
{\cal L}_{\rm soft}&=&-\left(\,
 A^u_{ij}H_2\tilde{q}_i\tilde{u}_j+A^d_{ij}H_1\tilde{q}_i\tilde{d}_j
+A^\ell_{ij}H_1\tilde{\ell}_i\tilde{e}_j
\right.\nonumber \\
&&\quad\left.+\,m^{2(\tilde{q})}_{i\bar{j}}\tilde{q}_i
\tilde{q}^*_j
+m^{2(\tilde{u})}_{i\bar{j}}\tilde{u}_i
\tilde{u}_j^*
+m^{2(\tilde{d})}_{i\bar{j}}\tilde{d}_i
\tilde{d}_j^*\right.
\nonumber \\
&&\quad\left.+\,m^{2(\tilde{\ell})}_{i\bar{j}}\tilde{\ell}_i
\tilde{\ell}^*_j
+m^{2(\tilde{e})}_{i\bar{j}}\tilde{e}_i
\tilde{e}_j^*\,\right).
\nonumber
\eea
There can be several different choices of $N_I\equiv
N(Q_I)$ which would yield the observed quark/lepton masses and mixing
angles \cite{choi3}. Here we will consider one example:
\bea
&&
N(q_i)=(-3,-2,1),\nonumber \\
&& N(u_i)=(-5,-2,1),
\nonumber \\
&&
N(d_i)=(-3,-2,-2),
\nonumber \\
&&N(\ell_i)=(-5,-2,-1),
\nonumber \\
&&N(e_i)=(-2,-2,-1).
\eea
These values of $N(Q_I)$  give the following forms of
Yukawa coupling matrices
\bea
\label{yukawamatrix}
y^u_{ij}&\,=\,& \pmatrix{ \epsilon^8\lambda^u_{11} & \epsilon^5
\lambda^u_{12} & \epsilon^3\lambda^u_{13} \cr
\epsilon^7\lambda^u_{21} & \epsilon^4\lambda^u_{22} & 
\epsilon^2\lambda^u_{23} \cr
\epsilon^5\lambda^u_{31} & \epsilon^2\lambda^u_{32} & 
\lambda^u_{33}}\,,
\nonumber \\
y^d_{ij}&\,=\,& \pmatrix{\epsilon^6\lambda^d_{11} & \epsilon^5\lambda^d_{12}
 &\epsilon^5\lambda^d_{13} \cr
\epsilon^5\lambda^d_{21} & \epsilon^4\lambda^d_{22} & \epsilon^4\lambda^d_{23}
 \cr
\epsilon^3\lambda^d_{31} & \epsilon^2\lambda^d_{32} &
\epsilon^2\lambda^d_{33}}\,,
\nonumber \\
y^{\ell}_{ij}&\,=\,&\pmatrix{
\epsilon^7\lambda^\ell_{11}&\epsilon^7\lambda^\ell_{12}&
\epsilon^6\lambda^\ell_{13}\cr
\epsilon^4\lambda^\ell_{21}&\epsilon^4\lambda^\ell_{22}&
\epsilon^3\lambda^\ell_{23}\cr
\epsilon^3\lambda^\ell_{31}&\epsilon^3\lambda^\ell_{32}&
\epsilon^2\lambda^\ell_{33}},
\eea
where $\lambda^{u,d,\ell}_{ij}$ are determined by the
brane Yukawa couplings $\lambda_{IJ}={\cal O}(\sqrt{\pi R\Lambda})$
and $N_I\ln(1/\epsilon)={\cal O}(m_I\pi R)$ as
$$\lambda^{u,d,\ell}_{ij}
={\cal O}\left(\frac{
\lambda_{IJ}\sqrt{N_IN_J}\ln(1/\epsilon)}{\pi R\Lambda}\right)
={\cal O}\left(\frac{m_I\pi R}{\sqrt{\pi R\Lambda}}\right)
={\cal O}(1).
$$
The soft parameters renormalized at $M_{KK}$ are determined to 
be
\bea
\label{softparameter}
\frac{A^u_{ij}}{y^u_{ij}}\,&=&\,
2M_{1/2}\ln 5\pmatrix{ 8 & 5 & 3\cr
7 & 4 & 2\cr
5 & 2 & 2\epsilon^2}\,,
\nonumber \\
\frac{A^d_{ij}}{y^d_{ij}}\,&=&\,
2M_{1/2}\ln 5\pmatrix{ 6& 5 & 5\cr
5& 4 & 4\cr
3& 2 & 2}\,,
\nonumber \\
\frac{A^\ell_{ij}}{y^\ell_{ij}}\,&=&\,
2M_{1/2}\ln 5\pmatrix{7&7&6\cr
4&4&3\cr
3&3&2}\,,
\eea
\bea
&&m^{2(\tilde{q})}_{i\bar{j}}=
(2\ln 5)^2|M_{1/2}|^2\,\pmatrix{
9\epsilon^6 &0&0\cr
0& 4\epsilon^4&0\cr
0&0& \epsilon^2}\,,\nonumber \\
&&\quad\quad\approx
|M_{1/2}|^2\,\pmatrix{
6\times 10^{-3}&0&0\cr
0& 6\times 10^{-2}&0\cr
0&0& 0.4}\,,
\nonumber \\
&&m^{2(\tilde{u})}_{i\bar{j}}=
(2\ln 5)^2|M_{1/2}|^2\,\pmatrix{25\epsilon^{10}&0&0\cr
0& 4\epsilon^4&0\cr
0&0& \epsilon^2}\,,
\nonumber \\
&&\quad\quad\approx |M_{1/2}|^2\,\pmatrix{
3\times 10^{-5}&0&0\cr
0& 6\times 10^{-2}&0\cr
0&0& 0.4}\,,
\nonumber \\
&&m^{2(\tilde{d})}_{i\bar{j}}=
(2\ln 5)^2|M_{1/2}|^2\,\pmatrix{9\epsilon^6&0&\cr
0&4\epsilon^4&0\cr
0&0&4\epsilon^4}\,,
\nonumber \\
&&\quad\quad\approx
|M_{1/2}|^2\,\pmatrix{6\times 10^{-3}&0&0\cr
0& 6\times 10^{-2}&0\cr
0&0&6\times 10^{-2}}\,,
\nonumber \\
&&m^{2(\tilde{\ell})}_{i\bar{j}}=
(2\ln 5)^2|M_{1/2}|^2\pmatrix{25\epsilon^{10} &0&0\cr
0&4\epsilon^4&0\cr
0&0&\epsilon^2}\,,\nonumber \\
&&\quad\quad\approx \left| M_{1/2} \right|^2 \, 
\pmatrix{ 
  3 \times 10^{-5} & 0 & 0 \cr
  0 & 6 \times 10^{-2} & 0 \cr 
  0 & 0 & 0.4 }\,,
\nonumber \\
&&m^{2(\tilde{e})}_{i\bar{j}}=
(2\ln 5)^2|M_{1/2}|^2\pmatrix{4\epsilon^4&0&0\cr
0&4\epsilon^4&0\cr
0&0&\epsilon^2}\,,
\nonumber \\
&&\quad\quad\approx \left| M_{1/2} \right|^2 \, 
\pmatrix{
  6 \times 10^{-2} & 0 &0 \cr
  0 & 6\times 10^{-2} & 0 \cr 
  0 & 0 & 0.4 }\,,
\nonumber 
\eea
where $M_{1/2}=F^T/2R$ denotes the universal gaugino
mass at $M_{KK}$.

The Yukawa coupling matrices of
(\ref{yukawamatrix})  produce
well the observed quark/lepton masses and also
the quark mixing angles for a reasonable range
of $\lambda^{u,d,\ell}_{ij}$.
The soft parameters of (\ref{softparameter})
are highly {\it flavor-dependent}, but appropriately suppressed
in correlation with Yukawa couplings.
After taking into account the renormalization group evolution
from $M_{KK}$  to the weak scale $M_W$,
the flavor-violations from the squark and slepton
masses at $M_{KK}$
can pass the known phenomenological constraints
for a reasonable range of parameters \cite{choi2}.
(Here we assume $M_{KK}$ is close to the unification scale
$2\times 10^{16}$ GeV.)
It is still true that the squark and slepton masses
induced by the SS boundary condition at $M_{KK}$
can lead to interesting flavor-changing
new physics signals at the weak scale,
which may be able to be observed in future experiments.
As for the squark $A$-parameters
$A^u_{ij}$ and $A^d_{ij}$, one arrives at a similar conclusion.

However the slepton $A$-parameter $A^{\ell}_{ij}$
can yield a too rapid 
 $\mu\rightarrow e\gamma$ unless $\lambda^{\ell}_{12,21}$
are appropriately tuned.
More explicitly,
to satisfy the experimental bound on $\mbox{Br}(\mu\rightarrow
e\gamma)$, one needs \cite{choi,choi2}
\bea
&&\lambda^\ell_{12} \lesssim 5\times 10^{-2}\left(\frac{
M_{1/2}}{500 \,\,\mbox{GeV}}\right)^2,
\nonumber \\
&&\lambda^\ell_{21}\lesssim 10^{-2}\left(
\frac{M_{1/2}}{500\,\,\mbox{GeV}}\right)^2.
\eea
This suggests that the holomorphic brane Yukawa
couplings of leptons at $y=0$, i.e. $\lambda_{IJK}$ in 
(\ref{braneaction}), 
conserve the lepton flavor $L_e$ or $L_\mu$,
for instance $L_e-L_\tau$, which would give
flavor-diagonal 
$$\lambda^{\ell}_{ij}=\lambda^{\ell}_i\delta_{ij}.
$$
Still one can achieve 
large lepton flavor-mixing in the neutrino mass matrix by
introducing gauge-singlet bulk neutrinos $N_i$ which
have {\it flavor diagonal} Yukawa couplings
$$\delta(y)\kappa_iH_2\ell_iN_i$$
at $y=0$ and 
{\it flavor non-diagonal}
Majorana mass masses  
$$\delta(y-\pi)M_{ij}N_iN_j$$
at $y=\pi$.
One can then adjust the kink masses of $N_i$ to make
the light neutrino mass matrix induced by the seesaw mechanism 
to take a form which can explain the observed large neutrino mixings
\cite{choi2}.

\medskip
\section{conclusion}
Imposing the Sherk-Schwarz boundary condition is an attractive
way to break SUSY in theories with compact
extra dimension. It can be interpreted as a spontaneous
SUSY breaking by the auxiliary component of
higher dimensional SUGRA multiplet.
Another attractive possibility associated with
extra dimension is the quasi-localization 
of matter fields which would generate hierarchical
Yukawa couplings in a natural manner.
In this talk, we discussed the flavor structure of soft SUSY breaking
parameters induced by the SS boundary condition
for quasi-localized quark/lepton superfields.
The resulting
squark/slepton masses and trilinear couplings
at the compactification scale 
are highly {\it flavor-dependent}, but appropriately suppressed
in correlation with Yukawa couplings.
Those flavor violations  
can give interesting phenomenological consequences
at low energies as well as constrain
the mechanism of Yukawa coupling generation.

\bigskip
\noindent
{\bf Acknowledgments}

This work is supported by KRF PBRG 2002-070-C00022
and the Center for High Energy Physics of Kyungbook National University.


\bigskip


\begin{thebibliography}{99}

\bibitem{nilles}
H. P. Nilles, Phys. Rep. {\bf 150}, 1 (1984);
H. Haber and G. Kane, Phys. Rep. {\bf 117}, 75 (1985).

\bibitem{ss}
J.~Scherk and J.~H.~Schwarz,
Phys.\ Lett.\ B {\bf 82}, 60 (1979);
\ Nucl.\ phys.\ B {\bf 153}, 61 (1979).

\bibitem{ss1}
D.~Marti and A.~Pomarol,
\prd {\bf 64}, 105025 (2001);
D.~E.~Kaplan and N.~Weiner,
hep-ph/0108001; 
G. v. Gersdorff and M. Quiros,
\prd {\bf 65}, 064016 (2002).


\bibitem{yukawa}
N.~Arkani-Hamed and M.~Schmaltz,
\prd {\bf 61}, 033005 (2000);
E.~A.~Mirabelli and M.~Schmaltz,
\prd {\bf 61}, 113011 (2000);
Y.~Grossman and M.~Neubert,
Phys. Lett. B {\bf 474}, 361 (2000);
S.~J.~Huber and Q.~Shafi,
Phys. Lett. B {\bf 498}, 256 (2001);
D.~E.~Kaplan and T.~M.~Tait,
JHEP {\bf 0111}, 051 (2001);
M.~Kakizaki and M.~Yamaguchi,
hep-ph/0110266; 
A.~Hebecker and J.~March-Russell,
Phys. Lett. B {\bf 541}, 338 (2002);
K. Choi, I-W. Kim and W. Y. Song, hep-ph/0307365.





\bibitem{choi}
K. Choi, D. Y. Kim, I.-W. Kim and T. Kobayashi,
hep-ph/0305024.

\bibitem{choi2}
H. Abe, K. Choi, I-W. Kim and K. S. Jeong, in preparation.

\bibitem{5dsugra}
A.~Ceresole and G.~Dall'Agata,
Nucl. Phys. B {\bf 585}, 143 (2000).











\bibitem{Arkani-Hamed:2001tb} 
N.~Arkani-Hamed, T.~Gregoire and J.~Wacker,
JHEP {\bf 0203}, 055 (2002);
D.~Marti and A.~Pomarol,
Phys. Rev. D {\bf 64}, 105025 (2001);
A.~Hebecker,
Nucl. Phys. B {\bf 632}, 101 (2002);
K.~Choi, H.~D.~Kim and I.-W.~Kim,
JHEP {\bf 0211}, 033 (2002);
JHEP {\bf 0303}, 034 (2003).


\bibitem{radion}
Z.~Chacko and M.~A.~Luty,
JHEP {\bf 0105} (2001) 067;
G. v. Gersdorff, L. Pilo, M. Quiros,
D. A. J. Rayner and A. Riotto,
hep-ph/0305218;
G. v. Gersdorff, M. Quiros and A. Riotto,
hep-th/0310190.



\bibitem{soft}
A.~Brignole, L.~E.~Ibanez and C.~Munoz,
Nucl. Phys. B {\bf 422}, 125 (1994);
V.~S.~Kaplunovsky and J.~Louis,
Phys.\ Lett.\ B {\bf 306}, 269 (1993);
K.~Choi, J.~S.~Lee and C.~Munoz,
Phys.\ Rev.\ Lett.\  {\bf 80}, 3686 (1998).






\bibitem{choi3}
K.~Choi, E.~J.~Chun and H.~D.~Kim,
\plb {\bf 394}, 89 (1997).




\end{thebibliography}
\end{document}